\documentclass{svproc}
\usepackage{url}

\usepackage{graphicx}
\graphicspath{{./figures/}}
\usepackage{amsmath,amssymb,stmaryrd}
\usepackage[authoryear]{natbib}

\newcommand{\df}{\mathrm{d}}

\begin{document}
\mainmatter              
\title{Development of a Computationally Efficient Fabric Model for Optimization of Gripper Trajectories in Automated Composite Draping}
\titlerunning{Development of a Computationally Efficient Fabric Model}  
%
\author{Christian Krogh\inst{1} \and Johnny Jakobsen\inst{1} \and James A. Sherwood\inst{2}}
\authorrunning{Christian Krogh et al.} 
%
\tocauthor{Christian Krogh, Johnny Jakobsen, James A. Sherwood}
\institute{Department of Materials and Production, Aalborg University, Denmark,\\
\email{ck@mp.aau.dk}
\and
Department of Mechanical Engineering, University of Massachusetts, Lowell, USA}

\maketitle              

\begin{abstract} 
An automated prepreg fabric draping system is being developed which consists of an array of actuated grippers. It has the ability to pick up a fabric ply and place it onto a double-curved mold surface. A previous research effort based on a nonlinear Finite Element model showed that the movements of the grippers should be chosen carefully to avoid misplacement and induce of wrinkles in the draped configuration. Thus, the present study seeks to develop a computationally efficient model of the mechanical behavior of a fabric based on 2D catenaries which can be used for optimization of the gripper trajectories. The model includes bending stiffness, large deflections, large ply shear and a simple contact formulation. The model is found to be quick to evaluate and gives very reasonable predictions of the displacement field. 
\end{abstract}
In the aerospace industry, composite parts are often made from woven carbon fiber prepregs, i.e. plies of woven carbon fiber tows that are pre-impregnated with a partially cured resin. The draping process, i.e. placement of the fabric plies onto a mold surface prior to curing, is currently done manually which constitutes a significant expense. Therefore, the research project \textit{FlexDraper} seeks to develop an automatic draping solution. 

The FlexDraper tool consists of an array of actuated grippers mounted to an industrial robot. To drape the fabric plies, the movement or trajectories of each of the grippers must be determined apriori. It is essential that the draped ply matches a prescribed boundary and does not include any out-of-plane wrinkles or in-plane waves. The problem was previously studied using an advanced nonlinear, rate-dependent Finite Element (FE) model \citep{Krogh2017a,Krogh2017}. In those previous studies, it was found that the placement of the ply is highly dependent on the path taken by the grippers and that wrinkles can easily form during draping. It was concluded that a concerted effort should be put into determining the optimal gripper trajectories. While the FE model agrees well with experimental results, it is also computationally expensive. Hence, it is not suitable for completing optimization analyses of the gripper trajectories. Therefore a faster, approximate model must be developed.

Several research groups are working with automated composite draping as outlined in the review by \cite{Bjornsson2017}. In the work by \cite{Molfino2014}, a robot cell for handling 3D carbon fabric is described, and a \textit{handling strategy} is outlined which relies on a computer vision system for generating control strategies. \cite{Flixeder2016} successfully set up a force based 2D handling system for fabrics where the movements of the manipulators are generated online. They employ a catenary model as basis for the controller for real-time prediction of the ply. In the study by \cite{Eckardt2016a}, a catenary model was used to calculate the inclination of the rotatable grippers such that no unnecessary bending is introduced in the ply. They tested three different manually created \textit{draping strategies} on a single curved demonstrator part. The first two strategies resulted in wrinkling or bridging while the third strategy was successful. In the paper by \cite{Brinker2017}, \textit{Drape paths} are discussed. A kinematic draping simulation is performed in steps which enables the gripper points to be tracked and trajectories generated. Other models as basis for automated handling of fabrics were introduced by \cite{Newell1995} (large deflection beam model) and \cite{Lin2009} (large deflection shell model).  

The present study focuses on developing a model that can estimate the 3D displacement field of a piece of fabric including interaction with the mold. It is to be used for determining the optimal gripper trajectories during draping onto double-curved molds with the FlexDraper system. The paper is organized as follows: Section 1 elaborates the robot system and presents the requirements to the model, Section 2 introduces the catenary fabric model, and Section 3 describes how the model is solved by means of optimization. Finally, Section 4 illustrates the usefulness of the model with a numerical example.
   
%

\section{Automated Draping with the FlexDraper System}
\begin{figure}[h!]
	\centering
	\includegraphics[width=0.57\linewidth]{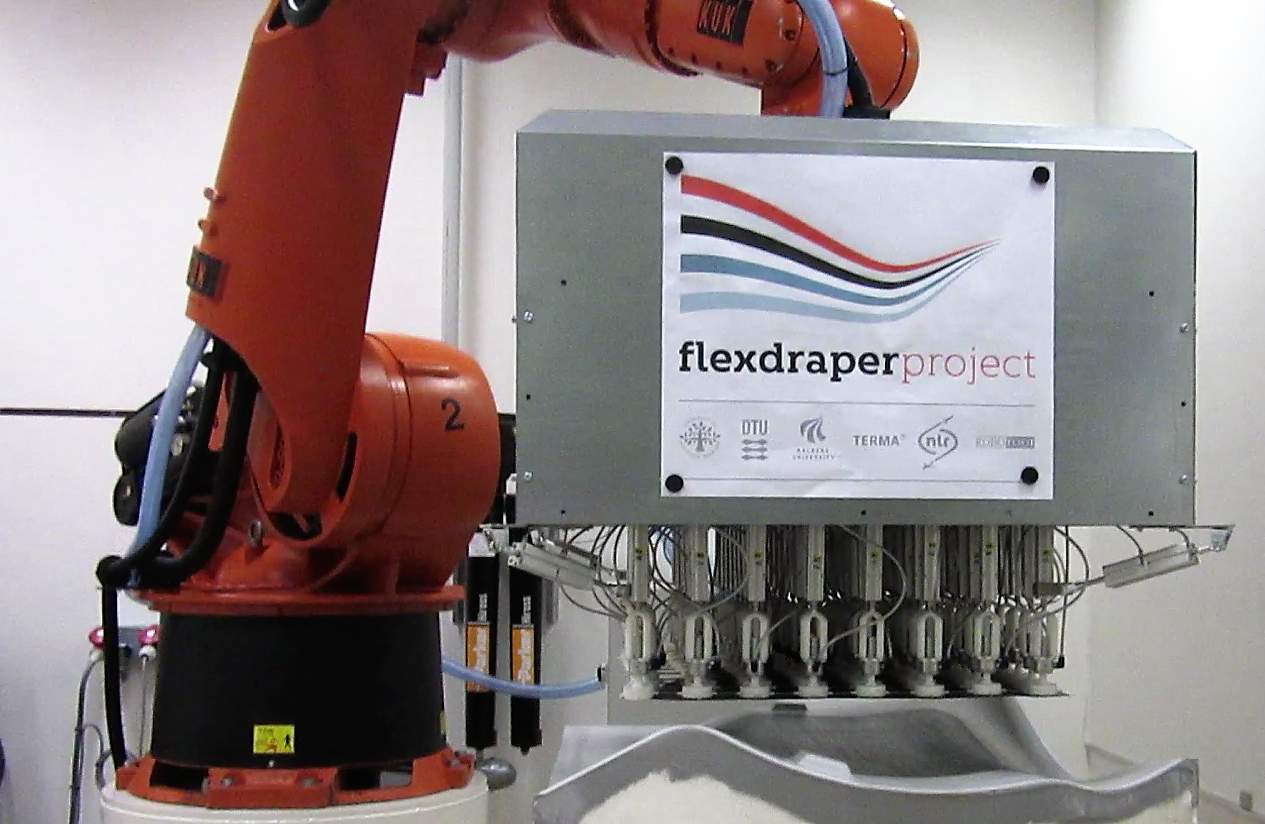}
	\caption{The FlexDraper robot cell with the ply on the grid of grippers.}
	\label{fig:robotcell}
\end{figure}

The FlexDraper robot system is depicted in Fig.~\ref{fig:robotcell}. All grippers are mounted in ball joints such that they can become tangent to the mold surface. In this picture, the ply has been picked up by the grid of grippers and is currently suspended over the mold. The next step is the actual draping, where the grippers move towards the mold surface. For the initially flat fabric to conform to a double curved surface, so-called \textit{fabric shearing} or trellising must take place. The fiber tows will rotate at the cross-over points in the weave such that the initially 90$^\circ$ angles between the two fiber directions will change.    

\subsection{Requirements to the Ply Model}
A woven carbon-fiber prepreg ply is comprised of two different materials, i.e. the carbon fibers and the resin, and spans different length scales. Thousands of approximately 6-$\mu$m diameter fibers are bundled in tows, which are woven into a fabric of unit cells with a size of approximately 5 mm$^2$. Yet, it often makes sense to consider the ply on a macroscopic level, which corresponds to the dimensions of the part to be manufactured. However, this approach also entails that the physics of the other two length scales must be considered as well. 

Experimental characterization of the ply has revealed that it exhibits nonlinear and rate-dependent stress-strain behavior in both the fiber directions and in shear. The rate-dependency can be accredited to the viscous behavior of the resin that also governs the frictional properties which come into effect upon contact with the mold. Also, the resin causes the fabric to have a significant bending stiffness. 

As previously described, the advanced Finite Element (FE) model is too computationally expensive for conducting optimization of the gripper trajectories.  The important task here is thus to identify which properties must be included in the approximate model. Based on the previous work with the prepreg material, the following were chosen as the requirements: 
\begin{itemize}
	\item Large deflections: equlibrium must refer to the deformed configuration.
	\item Bending stiffness: important for wrinkle size and shape prediction.
	\item Large ply shear: shear strains up to 30$^\circ$ are common in fabric draping.
	\item Mold contact: a simple formulation with an infinite coefficient of friction.
	\item Flexibility to change boundary conditions: the grippers can rotate due to the ball joints.
\end{itemize}
Based on the above requirements, it was chosen to consider a 2D catenary model with bending stiffness as the backbone of the model. As will become evident in the next section, it easily incorporates different boundary conditions and can be split in segments upon mold contact. Also, the shearing behavior can be decoupled from the fiber axial behavior.

\section{A Catenary Ply Model}
In the following, the 3D fabric catenary model is introduced. The governing differential equation is presented, the assembly of catenaries and shearing are described and finally the contact formulation is explained.   
\subsection{The Catenary Differential Equation} \label{sec:cat_eq}
\begin{figure}[b!]
	\centering
	\includegraphics[width=0.9\linewidth]{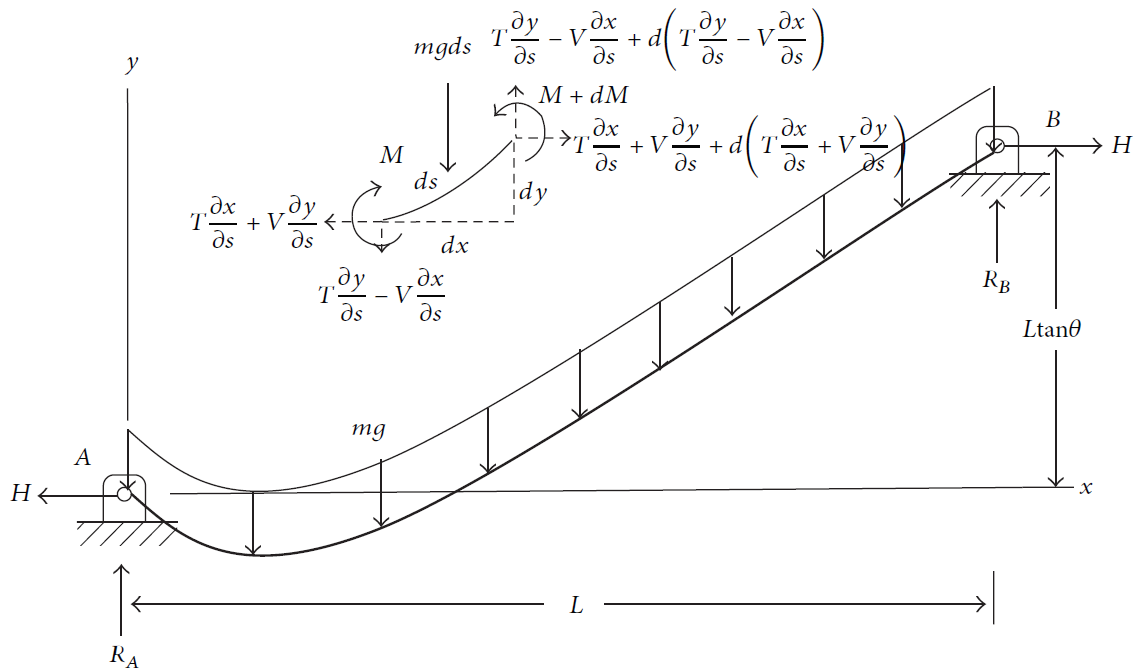}
	\caption{Free body diagram of catenary section with bending stiffness from \cite{Hsu2014}.}
	\label{fig:catenary_deriv}
\end{figure}
The differential equation of a catenary with bending stiffness is obtained using the free body diagram given in Fig.~\ref{fig:catenary_deriv} from \cite{Hsu2014}. By considering horizontal and vertical force equilibrium of an infinitesimal catenary element, the following can be obtained:
\begin{align}
	H \frac{\partial^2 y(x)}{\partial x^2} - \frac{\partial^2}{\partial x^2} \left[E I \frac{y''(x)}{\left(1+y'(x)^2\right)^{\tfrac{3}{2}}}\right] = m g \sqrt{1 + y'(x)^2}
\end{align}
where $H$, a constant, is the horizontal component of the catenary force $T$, $V$ and $M$ are the shear force and bending moment, respectively, $EI$ is the flexural rigidity and $mg$ is the weight per unit length. Unfortunately, no direct closed-form solution exists for this nonlinear differential equation. To simplify, a linear curvature definition is used and $EI$ is assumed to be independent of $x$. Also, small sag is assumed such that the gravitational load is uniformly distributed between the support points. Then, the following is obtained:
\begin{align}
	H y''(x) - EI y''''(x) = mg \sec(\theta) 
\end{align}
The small-sag assumption is usually considered valid when the sag to span ratio is less than 1/8. However, according to \cite{Irvine1981}, FE analyses have shown that the theory keeps its accuracy even with ratios as high as 1/4. 

The differential equation is solved with two different kinds of boundary conditions for each end depending on the gripper to which is attached. Either a zero moment condition, $y''(x) = 0$, such that the slope is free or a prescribed slope, $y'(x) = s_{pre}$. This use is elaborated in the next section.

The solution to the differential equation is the curve of a catenary with a length equal to what can be supported by the input reaction force. Hence, the next step is to apply length integration to find the reaction force $H$ that gives the right curve length. The length of a curve from point $a$ to point $b$ is analytically found as:
\begin{align}
	L = \int_a^b \sqrt{1+y'(x)^2} \df x
\end{align}
However, with the catenary solution, this integral cannot be evaluated. Another approach is to approximate the integral using e.g. Simpson's rule. While the integral can now be evaluated, the resulting expression cannot be solved for $H$. Therefore, it was chosen to find the reaction force numerically. The length is then evaluated as the sum of the Euclidean distances between all pairs of data points:
\begin{align}
	L = \sum_{i = 2}^{nPt} \sqrt{(x_i - x_{i-1})^2 + (y_i - y_{i-1})^2} \label{eq:length_calc_sum}
\end{align} 

\subsection{Systems of Catenaries} 
\begin{figure}[b!]
	\centering
	\includegraphics[width=0.68\linewidth]{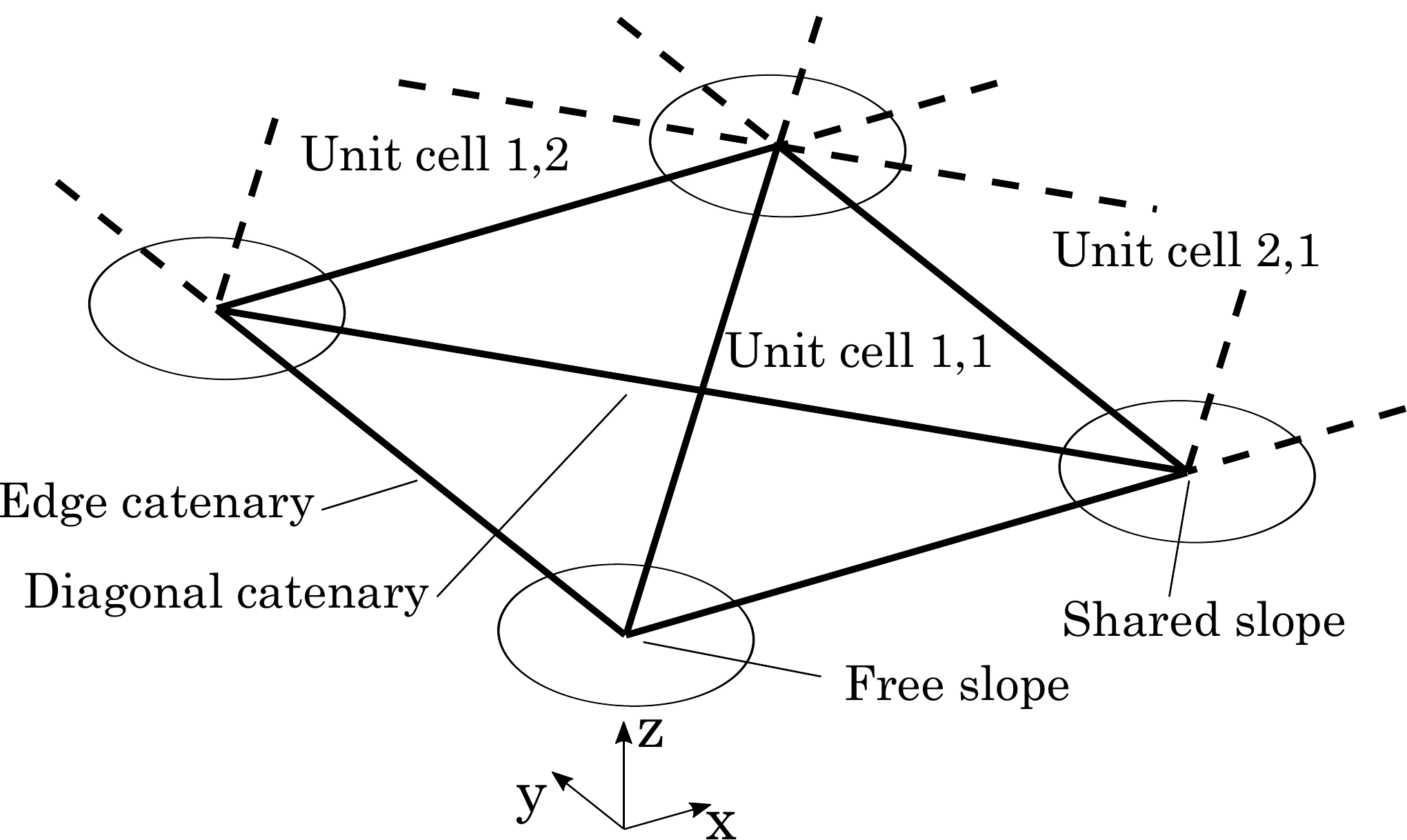}
	\caption{The assembly of catenaries into unit cells. Note also the concept of free and shared slopes.}
	\label{fig:catunitcell}
\end{figure}

To model a 3D fabric using 2D catenaries, unit cells are created in between the grippers as depicted in Fig.~\ref{fig:catunitcell}. 
Edge catenaries are suspended between adjacent grippers and diagonal catenaries are suspended diagonally across the unit cells. The former will account for the fiber direction behavior while the latter will account for shear. The $z$ component of position is the catenary solution whereas the $x$ and $y$ components are linear interpolations between the grippers. In the case where the diagonals in a unit cell do not intersect, a post-processing approach is employed: the lowest diagonal is split in two half segments and re-evaluated with the new center point being coincident with the highest diagonal. 

The rotations are determined according to the following rules:
\begin{itemize}
	\item Free-edge catenary ends are assigned a free slope.
	\item Colinear edge catenaries with a shared gripper are assigned the same slope which is determined during solution.
	\item Diagonals are assigned a slope which is calculated based on the surrounding edge catenaries. 
\end{itemize}
%

\subsection{Shearing}
\begin{figure}[b!]
	\centering
	\includegraphics[width=0.65\linewidth]{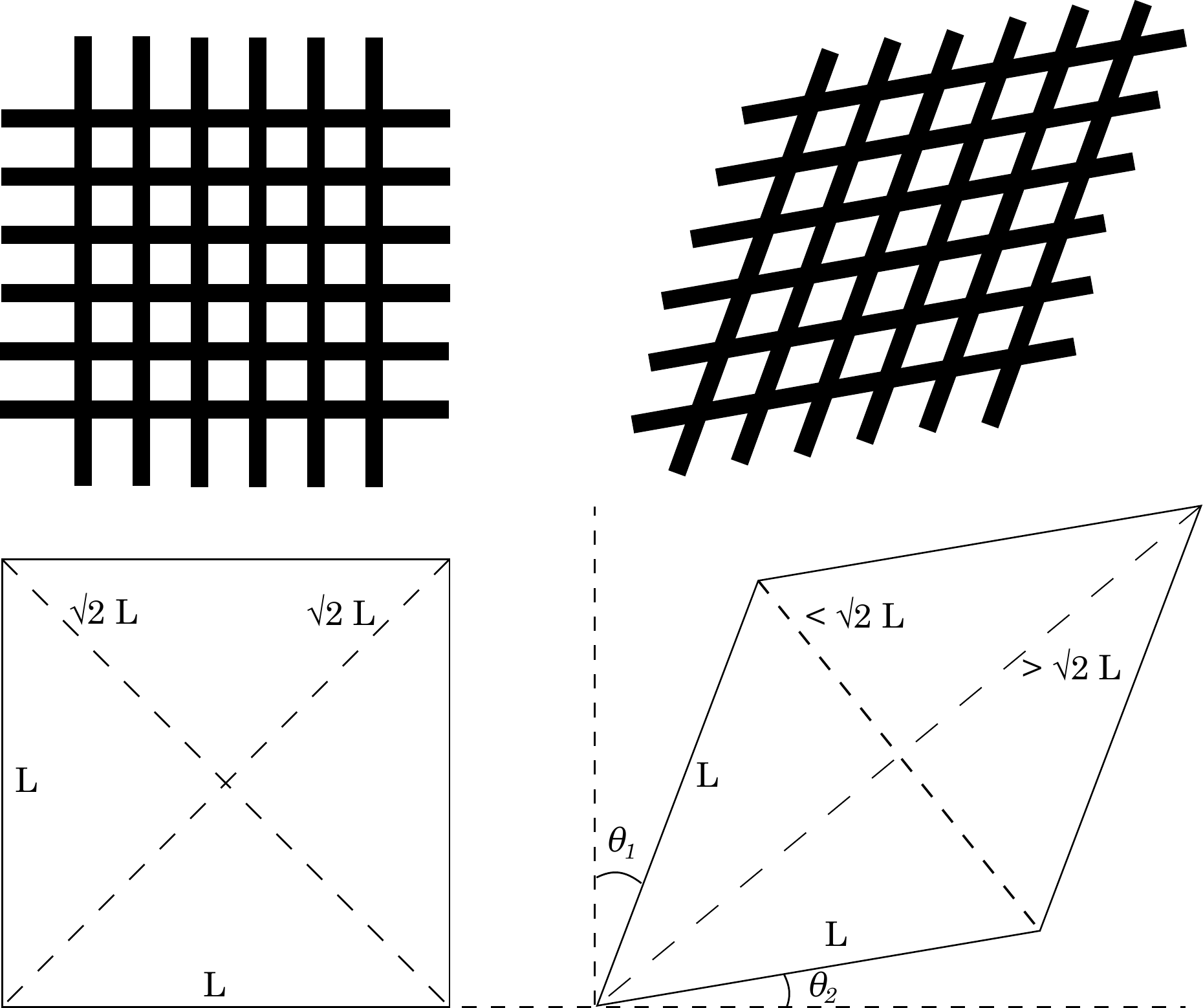}
	\caption{The kinematics of a gripper unit cell undergoing shear deformation.}
	\label{fig:shearkinematics}
\end{figure}

As previously described, the fabric must shear to conform to a double-curved surface. When the diagonal between two grippers exceeds the prescribed length of the pertaining diagonal catenary, the unit cell is considered to be shearing. Only one diagonal can shear in each unit cell. When shearing, the unit cells deform like parallelograms. Thus, if one diagonal extends, the other must become shorter as sketched in Fig.~\ref{fig:shearkinematics}. Notice that the diagonal catenaries do not represent actual fiber tows.
When a diagonal is shearing, a catenary solution is not desired. Instead, the $z$ component of position is taken as a linear interpolation, i.e. the shearing diagonal will be a straight line across the unit cell. The reaction force is then found using an empirical model as determined from experimental shear-test data.

Another thing to note about a shearing diagonal in the model is that it will govern the slopes of all other surrounding catenaries. That is, the edge catenaries sharing a gripper with a shearing diagonal will have the same slope as the shearing diagonal.    

\subsection{Contact Formulation}
\begin{figure}[htb]
	\centering
	\includegraphics[width=1.0\linewidth]{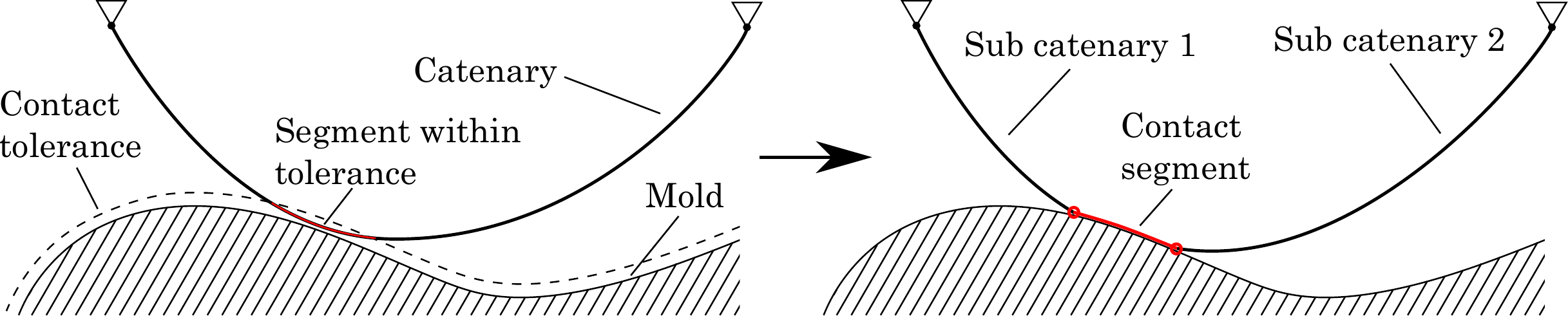}
	\caption{Evaluating contact for the model. Left: A segment of the catenary is within the contact tolerance of the mold. Right: the contact segment is fixed and two independent catenaries are created.}
	\label{fig:catcontact}
\end{figure}

For the mold-ply contact, a simple infinite friction formulation is used. Basically, when a segment of the catenary is within a given tolerance of the mold, that segment is considered in contact with the mold. Hereafter, it will remain on the mold and cannot move in any direction. This situation is shown in Fig.~\ref{fig:catcontact}.

The non-contacting parts of the catenary are now made into independent catenaries. Their length is calculated based on the length of the contact segment. For simplicity, only one contact segment can exist between two grippers. If multiple contact segments existed, it would likely be an indication of ply bridging.  

\section{Solution Method} \label{sec:solution}
With the catenaries arranged in unit cells, the unknown parameters, i.e. horizontal reaction forces and shared slopes must be determined such that the right lengths are obtained. It was decided to formulate a nonlinear constrained optimization problem to be solved using Sequential Quadratic Programming (SQP):
\begin{alignat}{3}
\underset{\mathbf{H,s}}{\text{minimize}}&~~\sqrt{\mathbf{H}\,\mathbf{H}^T}&&\notag\\
\text{subject to}&~~L_i = L_{set,i} ~~ \forall i \qquad&&,~i=1,...,nCat\\
&~~s_{min} \leq s_j \leq s_{max} ~~ \forall j \qquad&&,~j=1,...,nSharSlope \notag
\end{alignat}
The design variables are the horizontal reaction forces $\mathbf{H} = \{H_1, H_2,...,H_{nCat}\}$ and the shared slopes $\mathbf{s} = \{s_1, s_2,...,s_{nSharSlope}\}$. The objective is to minimize the reaction forces, mathematically formulated as the 2-norm of the vector. The equality constraints dictate that the length of each catenary must be equal to the set length. Also, lower and upper bounds are imposed on the slopes which correspond to the maximum rotation of the grippers in the ball joints (40$^\circ$). The catenary fabric model is implemented in MATLAB and solved using the built-in \textit{fmincon} function.
\subsection{Design Sensitivity Analysis}
Gradients of the objective function and the nonlinear length constraints are needed for the SQP solver. Because the objective function is an analytical function of the components of \textbf{H}, an analytical gradient is readily available. The details are not presented here.

For the nonlinear constraints, the issues discussed in Sec.~\ref{sec:cat_eq} regarding length integration also inhibits the calculation of analytical gradients. However, it is still a substantial improvement to consider \textit{semi-analytical} gradients on the catenary level compared to an Overall Finite Difference (OFD) calculation using the constraint function. This statement is especially true because each catenary length is only a function of the corresponding catenary reaction force and endpoint slopes. Thereby, many sensitivities are equal to zero. 

In general, the analytical catenary solution $z$ is a function of $x$ and $y$, the horizontal reaction force $H$, the two slopes $s_1$ and $s_2$ and the material parameters here denoted \textit{Mat} for convenience:
\begin{align}
	z = z(x,y,H,s_1,s_2,\mathit{Mat})
\end{align}
At any given design point where the sensitivity is sought, all of these quantities are known. Consider for example the sensitivity of a length constraint of a catenary to the corresponding horizontal reaction force. The calculations are carried out according to the following procedure:
\begin{enumerate}
\item Evaluate catenary solution at design point: $z_H = z(x,y,H,s_1,s_2,\mathit{Mat})$ 
\item Perturb $H$ and evaluate again: $z_{H + \Delta H} = z(x,y,H + \Delta H,s_1,s_2,\mathit{Mat})$  
\item Calculate lengths c.f. Eq.~\eqref{eq:length_calc_sum}: $L_H,L_{H + \Delta H}$
\item Evaluate finite difference approximation: $\frac{\partial L}{\partial H} \approx \frac{L_{H + \Delta H}-L_H}{\Delta H} $
\end{enumerate}  
The same principle can be applied for the sensitivities of the length constraint to the slope design variables.       

\section{Numerical Example}
As a proof of concept, an example with a flat sheared mold, i.e. a generic parallelogram, and a grid of 4 $\times$ 2 grippers is presented. In place of the gripper trajectory optimization, the movement of each gripper will be controlled by a linear interpolation from the initial position to the target point on the mold. The target points correspond to the positions of the grippers in an ideal draped configuration. 

\begin{figure}[h!]
	\centering
	\includegraphics[width=1.0\linewidth]{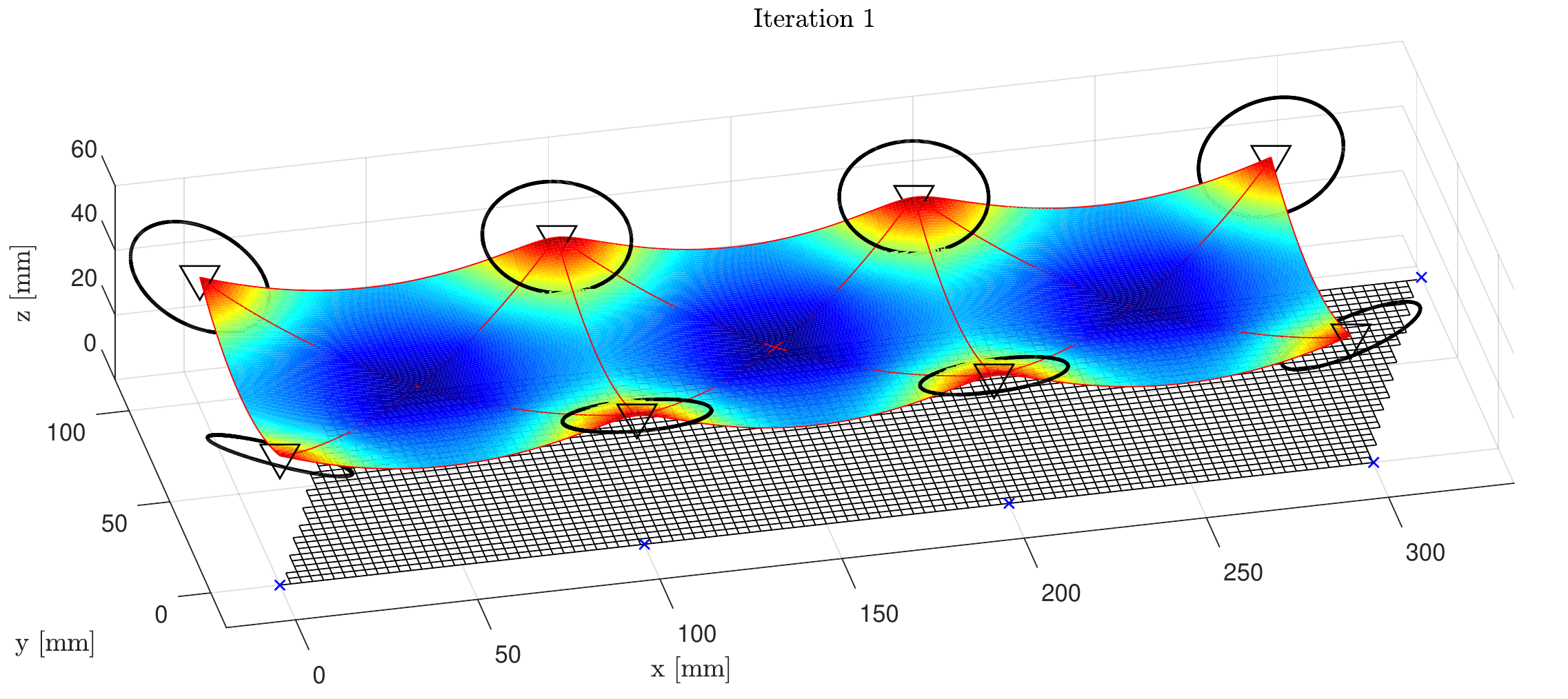}
	\includegraphics[width=1.0\linewidth]{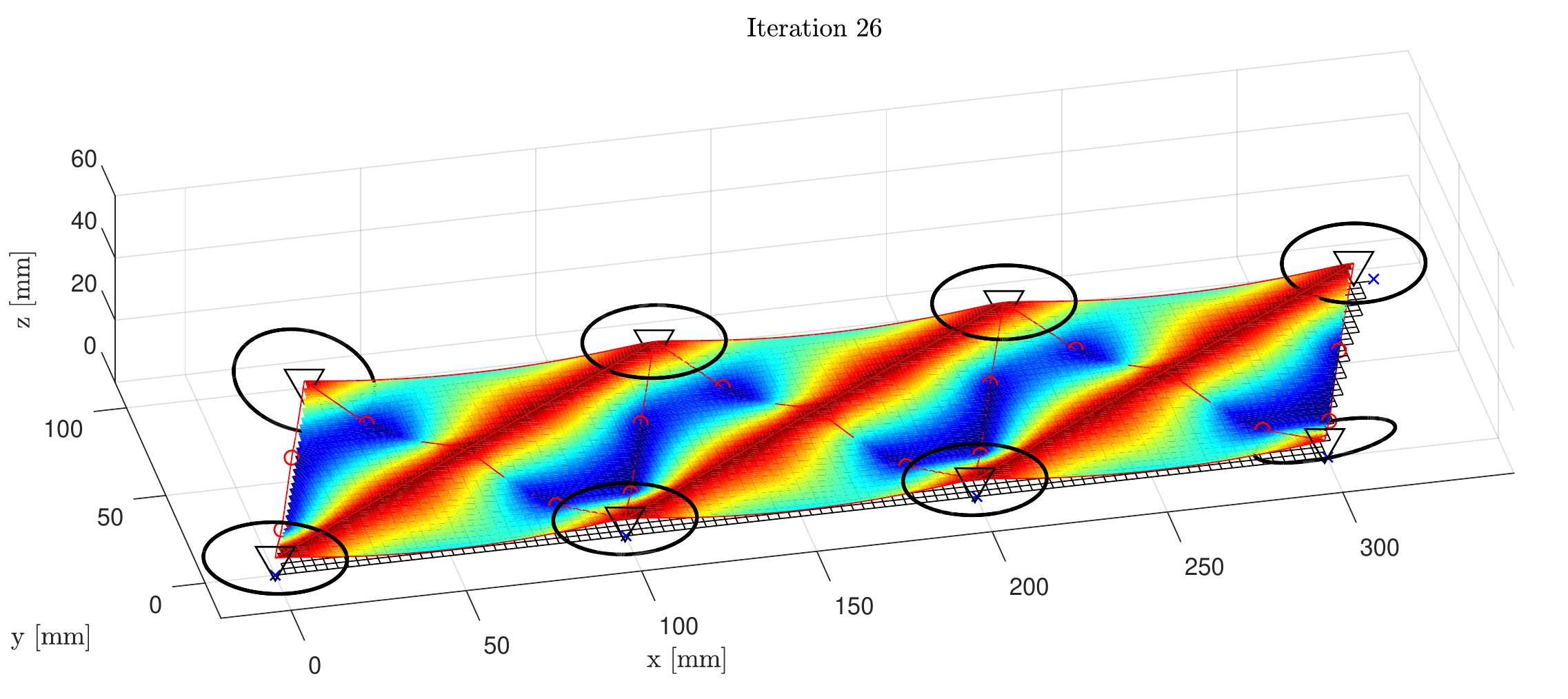}
	\includegraphics[width=1.0\linewidth]{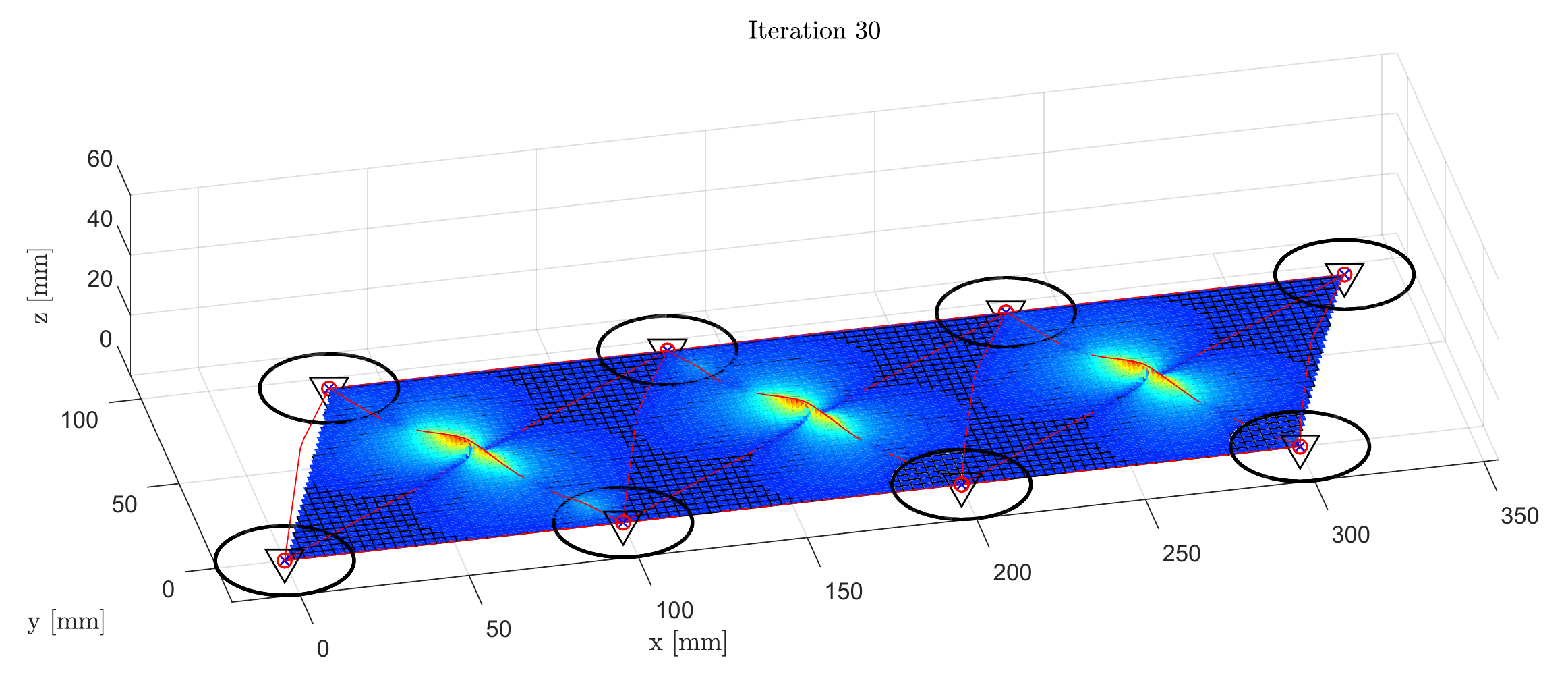}
	\caption{Different iterations in the draping of fabric onto flat sheared mold.}
	\label{fig:drape}
\end{figure}

Initially, the grippers start in a rectangular grid with a spacing of 98 mm. The length of the gripper unit cell is set to 100 mm, and therefore the model will predict some slack. From here, the grippers are moved to their target points on the mold in 30 iterations of which three are presented in Fig.~\ref{fig:drape}. Model input data are listed in Table~\ref{tab:modpar}.
\newpage
In Fig.~\ref{fig:drape} the catenaries are shown as red lines. For visualization purposes, a surface is interpolated inside the unit cells and colorized based on the $z$ coordinate. The grippers are also visualized as black circles. Notice, however, that they do not enter into the model itself but are created based on the catenary slopes. In each iteration, the grippers are moved and the state of contact is evaluated. Iteration 1 is the initial configuration. In iteration 26, the ply is shearing and has just made initial contact with the mold. The current contact points are indicated with red circles. Because of the sag, the ply makes contact with the mold before the grippers have reached the $x$ and $y$ coordinates of the mold target points. As a result, wrinkles are formed and remain in the draped configuration, i.e. iteration 30, due to the infinite friction assumption. Notice that the model does not give a physical representation of the wrinkles as the two diagonals in a unit cell are no longer coincident. This behavior can be attributed to the fact that the diagonals and the contact formulation are decoupled. For the present purpose, however, a good visual representation of the wrinkles is not needed. In fact, it suffices with an indication of whether the ply is wrinkling or not because the goal is a draping sequence without wrinkles. Consequently, a \textit{wrinkling criterion} could be implemented at a later stage.         

It is worth noting that the model does predict a realistic displacement field throughout the iterations. In terms of speed, the typical model evaluation time is 0.2 - 0.3 s on a standard PC. 
\begin{table}
\caption{Model parameters for the flat sheared mold draping example.}
\begin{center}
\begin{tabular}{p{0.18\textwidth}p{0.18\textwidth}p{0.18\textwidth}p{0.18\textwidth}p{0.18\textwidth}}
\hline
\rule{0pt}{12pt}$m$ & $g$ & $E$ & $I$ & Mold shear \\[2pt]
\hline\rule{0pt}{12pt}
0.3143 kg/m  & 9.8 m/s$^2$ & 1.0e8 Pa & 9.0e-14 m$^4$ & 20$^\circ$ \\[2pt]
\hline
\end{tabular}
\end{center}
\label{tab:modpar}
\end{table}   
\section*{Conclusion}
A computationally efficient model of a deformable piece of fabric has been developed. It is intended to be used with a robot system for automatic draping of fiber plies. The large deflection model consists of a system of catenaries suspended between the grippers of the robot tool. The model can estimate the 3D displacement field of the fabric and includes the bending stiffness of the ply, which is important for the shapes and sizes of wrinkles. The model decouples the shearing behavior from the fiber axial behavior. This decoupling enables the model to account for large ply shear which is necessary when draping fabric onto double curved surfaces. Also, a simple contact formulation is implemented such that the mold-ply contact can be captured. Contact is established when a segment of a catenary is within a given tolerance of the mold surface. 

The model is solved using a  nonlinear constrained optimization scheme which minimizes the reaction forces and constrains the lengths of the catenaries to their prescribed values. The design variables are the reaction forces and slopes shared between catenaries. Analytical gradients of the objective function and semi-analytical gradients of the constraints are provided for increased speed. 

The numerical example showed the ability of the fabric model to capture both sagging and shearing of the ply as well as the mold-ply contact. Although no trajectory optimization was employed, the example highlighted the difficulties with feasible draping sequences. Because the ply made contact with the mold at inconvenient locations, the model indicated wrinkles in the draped configuration. This phenomenon is exactly the motivation for employing optimization for determining the gripper trajectories. 

\section*{Future Work}
The gripper trajectory optimization problem must be formulated and implemented in the program along with the model. The approach is to divide the gripper movement into steps, where in each step an optimization problem is solved with move limits imposed on the grippers. Before mold contact, the objective is - in addition to minimizing the mold-ply distance - to position the ply such that it will match the prescribed boundary at the end of the placement process. After mold contact, the objective is to minimize the difference between the slopes of the ply and the mold at the contact point. In this manner, the ply will be rolled onto the mold such that wrinkling and bridging will be mitigated. Ultimately, the generated gripper movements are used in the FE model and compared to experimental results 

\section*{Acknowledgements}
This research presented in this paper is part of the project FlexDraper - An Intelligent Robot-Vision System for Draping Fiber Plies sponsored by the Innovation Fund Denmark, Grant no. 5163-00003B. This support is gratefully acknowledged. The authors
also thank all partners within the project.


%

%
\bibliography{literature}

\end{document}